\documentclass{cup-hpl}

\begin{document}

\shorttitle{A thick-target probe for ultra-low-energy resonances}
\shortauthor{X.-Y. Luo et al.}

\title{Bypassing the Nanoampere Limit: A Thick-Target Methodology for Ultra-Low-Energy Resonances in the $^{11}$B(p,$\alpha$)2$\alpha$ Reaction}

\author[1]{Xu-Yang Luo}
\author[1]{Wen-Qing Wei\corresp{W. Q. Wei and Y. T. Zhao, School of Physics, Xi'an Jiaotong University, Xi'an 710049, Shaanxi, China.
\email{wenqingwzx@xjtu.edu.cn (W. Q. Wei); zhaoyongtao@xjtu.edu.cn (Y. T. Zhao)}}}
\author[1]{Shi-zheng Zhang}
\author[1]{Wen Jiang}
\author[1]{Bu-bo Ma}
\author[1]{Zi-qian Zhao}
\author[2]{Zhao Wang}
\author[2]{Guo-Dong Wang}
\author[2]{Lu-Lin Shi}
\author[2]{Rui Cheng}
\author[2]{Yan-Shi Zhang}
\author[3]{Di Luo}
\author[3]{Xiao Xiao}
\author[3]{Guan-Chao Zhao}
\author[3]{Bing Liu}
\author[1,3]{Dieter H.H. Hoffmann}
\author[1]{Jie-Ru Ren}
\author[1]{Yong-Tao Zhao}

\address[1]{Ministry of Education Key Laboratory for Nonequilibrium Synthesis and Modulation of Condensed Matter, Shaanxi Province Key Laboratory of Quantum Information and Quantum Optoelectronic Devices, School of Physics, Xi'an Jiaotong University, Xi'an 710049, Shaanxi, China}
\address[2]{Institute of Modern Physics, Chinese Academy of Sciences, Lanzhou 730000, Gansu, China}
\address[3]{Beijing ENN Fusion Energy Science and Technology Co., Ltd., Beijing 101111, Beijing, China}

\begin{abstract}
Precise measurements of the low-energy cross section for the $^{11}$B(p,$\alpha$)2$\alpha$ reaction is critical for nuclear astrophysics and fusion energy research, yet such measurements are experimentally demanding due to the reaction’s extremely small cross section. Conventional thin-target experiments are fundamentally limited to nanoampere beam currents, which restrict count rates to impractical levels at ultra-low proton energies. To overcome this limitation, we report a correction-free thick-target methodology capable of operating at beam currents of 20-60 microamperes, corresponding to a gain of over three orders of magnitude. Monte Carlo simulations of the full reaction kinematics are employed to determine the absolute reaction yield, eliminating the requirement for empirical normalization factors. At the well-known 148\,keV (c.m.) resonance peak, our extracted resonance energy ($E_{\mathrm{r}} = 148.0$\,keV) and level width ($\Gamma_{\mathrm{cm}} = 5.3$\,keV) are demonstrated in good agreement with published data. A scheme via comparing the measured reaction probability with a theoretical extrapolation result is furtherly presented to probe low-energy resonances below 100\,keV region. Preliminary measurements at proton energies of 75\,keV and 100\,keV reveal no evidence for additional low-energy resonances. This work provides a robust new methodology to measure cross section
and resonance parameters for the p$^{11}$B reaction and offers a highly efficient tool for systematic resonance searches in the ultra-low-energy regime.
\end{abstract}


\maketitle

\section{Introduction}\label{sec1}

The $^{11}$B(p,$\alpha$)2$\alpha$ reaction has attracted considerable interest in
advanced fusion energy~\cite{Moreau_1977,app12031444,Valeria2023,Batani_Margarone_Belloni_2023,PhysRevE.106.055215,Ogawa_2024,Magee2023,li2024proton},
laser nuclear physics~\cite{Labaune2013FusionRI,giuffrida_high-current_2020,bonvalet_energetic_2021,picciotto_boron-proton_2014,luo2026gated,sciscio2025laser},
and cancer therapy such as proton-boron capture therapy (PBCT)~\cite{Jelinek2024,Yoon2014,Tran2023CurrentSA}, owing to its
neutron-free products, high energy release (8.7 MeV per reaction), and abundant
boron resources~\cite{Mckenzie2023}. The total cross section and resonance behavior
of this reaction are essential for evaluating fusion reactivity in both magnetic
and inertial confinement plasmas~\cite{sikora_new_2016}. They also provide key information for understanding
the excited state structure of the compound nucleus $^{12}$C~\cite{kelley_energy_2017,Laursen2016},
nuclear reaction mechanisms~\cite{buck_alpha-cluster_nodate,stave_understanding_2011}, and the origin of light elements in astrophysics~\cite{Boesgaard_2005}.
Precise measurement of the $^{11}$B(p,$\alpha$)2$\alpha$ cross section at low energies,
particularly the resonance parameters of the first resonance peak,
has therefore remained a significant topic in nuclear physics experiments~\cite{Tentori_2023}.
However, due to Coulomb repulsion, the
cross section drops exponentially with decreasing energy in the low-energy region, making direct measurements extraordinarily difficult~\cite{iliadis_2007}.

Beyond the intrinsic suppression of the cross section, a key challenge in experimental determination of the $^{11}$B(p,$\alpha$)2$\alpha$ cross section is the normalization of the absolute reaction yield. The p-$^{11}$B reaction proceeds via two main channels: the ground-state channel ($\alpha_0$) and the first excited state channel ($\alpha_1$). Each channel involves a primary two-body reaction
$^{11}$B(p,$\alpha$)$^{8}$Be followed by the prompt decay of $^{8}$Be into two $\alpha$ particles. This three-body final state complicates $\alpha$-particle kinematics and detection. Since no ideal purely discrete $\alpha$ energies appear, the three $\alpha$ particles partition the available energy, yielding a overall spectrum with a continuous distribution determined by their kinematic correlations. This makes particle identification, spectrum unfolding and absolute yield extraction substantially more involved than for two-body nuclear reactions. While early measurements spanning the
1930s-1960s established the basic features of this
reaction~\cite{oliphant_experiemnts_1933,dee_disintergratin_1936,beckman_excitation_1953,segel_states_1965},
the normalization issue became apparent in subsequent systematic studies. Davidson $\textit{et al.}$
(1979)~\cite{davidson_low_nodate} divided the total $\alpha$-particle yield by 3, assuming all three $\alpha$ particles from each reaction event were detected. Becker $\textit{et al.}$ (1987)~\cite{becker_low-energy_nodate} argued that their configuration captured
only two of the three particles and divided by 2, leading to a factor of 1.5 discrepancy. Munch and Fynbo~\cite{Munch2018ThePW} later resolved this ambiguity
using double sided silicon strip detectors (DSSDs) to reconstruct the full kinematics, demonstrating that division by 3
was valid. However, all of these approaches rely on an empirical division
factor to normalize the yield. Moreover, DSSD measurements require complex multi-detector arrays, making
systematic scans over a wide energy range cumbersome. These challenges become even more severe at low proton energies. When the proton energy drops below 100\,keV, the reaction cross section decreases sharply, and
the already low count rate falls below the experimentally feasible detection threshold. 

To overcome such low-count-rate limitations induced by tiny cross sections, low-energy reaction studies are also performed in deep underground laboratories, such as JUNA at the China Jinping Underground Laboratory~\cite{zhang2022measurement} and LUNA at Gran Sasso~\cite{BROGGINI201855}, where the cosmic-ray background is suppressed by kilometers of rock shielding. These facilities have extended direct measurements of several key reactions to unprecedentedly low energies. At these low energies, thin-target measurements face a fundamental limitation: the extremely small cross section demands high beam intensities to achieve adequate statistics, yet conventional thin targets are limited to nanoampere currents due to thermal load constraints. In contrast, thick targets can sustain
beam currents of tens of microamperes, directly compensating for
the low cross section. Target preparation is straightforward and requires no precise thickness control, while the method is less sensitive to beam-energy fluctuations. These characteristics make thick-target experiments particularly suitable for systematic low-energy cross-section measurements. However, extracting accurate cross sections and resonance parameters from thick-target yield data requires a reliable inversion methodology.

In this work, we developed such a method for the p$^{11}$B reaction to measure the cross section and resonance parameters. Monte Carlo simulations of the full reaction kinematics were used to directly determine the reaction yield, without relying on empirical division factors. This method was demonstrated on the first resonance peak by comparing the extracted resonance parameters with established values, showing good agreement between them. We further extend the approach into the ultra-low-energy regime ($<$ 100\,keV), where the high beam current of the thick-target method compensates for the extremely small cross section. By comparing measured thick-target reaction probability with extrapolations that assume no additional resonance, any systematic excess would indicate the presence of a low-energy resonance, providing an efficient probe for fast systematic searches.

This paper is organized as follows: the experimental setup together with target and detector parameters are described in Section 2; Main results and discussion, including the Monte Carlo methodology for cross-section extraction, the extraction of resonance parameters, and the application of this method in probing low-energy resonances are presented in Section 3; Finally, conclusions are summarized in Section 4.

\section{Experimental Setup}\label{sec2}

\begin{figure}[htbp]
\centering
\includegraphics[width=0.48\textwidth]{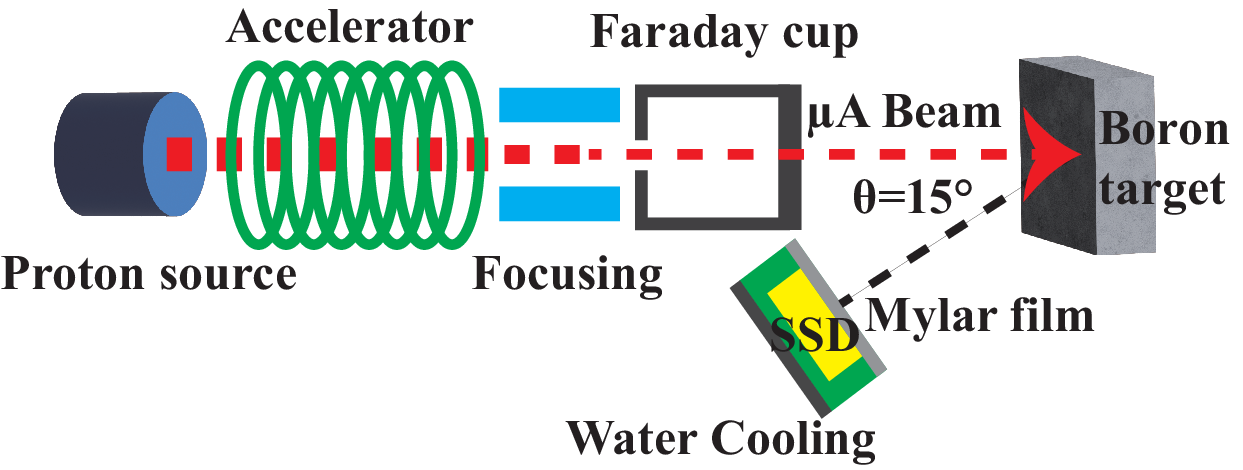}
\caption{Schematic of the experimental setup. An accelerated proton beam irradiated a thick solid boron target to trigger the p-$^{11}$B reaction in the vacuum chamber. The emitted $\alpha$ particles were detected by a silicon surface barrier detector (SSD) positioned at 45~cm away from the target front surface along $165^\circ$ relative to the target normal direction. A 4\,$\mu$m Mylar film was placed in front of the detector, and a Faraday cup was used to monitor the beam current at the end of the beamline. A water cooling system was integrated into the detector setup to mitigate the influence 
of ambient temperature fluctuations on the detector. }
\label{fig1}
\end{figure}

The experiments were carried out at the high-power proton linear accelerator facility at the Institute of Modern Physics, in Lanzhou. The accelerator delivers proton beams with energies from 40 to 240 keV and a beam spot radius of less than 10 mm, in both direct current (DC) and pulsed modes ~\cite{2026LuoDi}. The beam operated in the pulsed mode has a repetition rate of 1 Hz and a pulse width of 500 ms. This mode reduces the beam average power and prevents target heating, while provides sufficient statistics within each pulse. The schematic diagram of the experimental setup is shown 
in Fig.~\ref{fig1}. The proton beam produced from an ion source is accelerated through a 150~mm-length acceleration tube to the designated energy, and focused by quadrupole magnets, which is finally introduced into the vacuum chamber for the proton-boron reaction experiments. The beam current is measured by a stopping Faraday cup at the end of the beamline. Unlike conventional thin-target experiments, where the beam current is limited to the nanoampere range, the thick-target configuration of the present setup allows operation at tens of microamperes.

Thick solid boron targets with natural isotopic abundance ($^{11}$B $\approx$ 80.1\%) were used.
The target thickness was 3.3 mm, and the nominal density was 1.113 g/cm$^3$, which was sufficient to stop the incident protons completely. Because the target surface was exposed to air, carbon and oxygen contamination layers formed at both side surface 
readily. Rutherford backscattering spectrometry (RBS) with a 5.7 MeV $^{4}$He beam was employed to characterize the composition and areal density of these surface layers.
In the outermost layer (areal density $700 \times 10^{15}$ atoms/cm$^2$), RBS analysis
revealed carbon of $\sim$3.4 at.\%, oxygen of $\sim$3 at.\%, and a heavy impurity element with a mass number close to that of Mg (not confirmed as Mg) of $\sim$27 at.\%. The $^{11}$B content in this layer was correspondingly reduced to $\sim$67 at.\%. Below this layer, the $^{11}$B content increased to 82-89 at.\%, consistent with the natural isotopic abundance of the bulk target.
The energy loss of incident protons in the surface layer was calculated from the RBS
areal density using SRIM stopping cross-sections~\cite{ziegler_srim_2010}, giving a correction range from 8 keV to 15 keV depending on the incident energy. This correction was applied in the subsequent data analysis.

Due to the thick-target configration, the reaction product $\alpha$ particles emitted along the beam forward direction are stopped within the target material. Hence, $\alpha$ particles were detected at a beam backward angle of $165^\circ$ relative to the target normal direction, using a silicon surface barrier detector.
The detector had a diameter of 2 cm and was located approximately 45 cm from the target front surface, giving a solid angle of $\Delta\Omega \approx 1.55\times 10^{-3}$ sr (half-angle $\approx 1.3^\circ$). A 4 $\mu$m-thick Mylar film was placed in front of the detector to block low-energy stray particles, secondary electrons, and scatter light from the beam-target interaction. With a thickness of 300 $\mu$m, this detector provided sufficient stopping power to fully stop $\alpha$ particles and enable their complete detection. The detector was calibrated using a three-component standard $\alpha$ radioactive source, yielding the linear calibration function $E = 0.00185c + 0.98849$, where $c$ denotes the channel address of the detector, $E$ is the $\alpha$ particle energy \cite{2026LuoDi}.  Measurements were performed at proton energies from 156 to 220 keV, with fine steps of approximately 2 keV across the resonance region (156-170 keV) and coarser steps at higher energies. Additional measurements at 75 keV and 100 keV were carried out for the low-energy resonance search.

\section{Results and Discussion}\label{sec3}

\subsection{Thick-target yield and cross-section extraction}\label{subseci1}
\begin{figure}[htbp]
\centering
\includegraphics[width=0.48\textwidth]{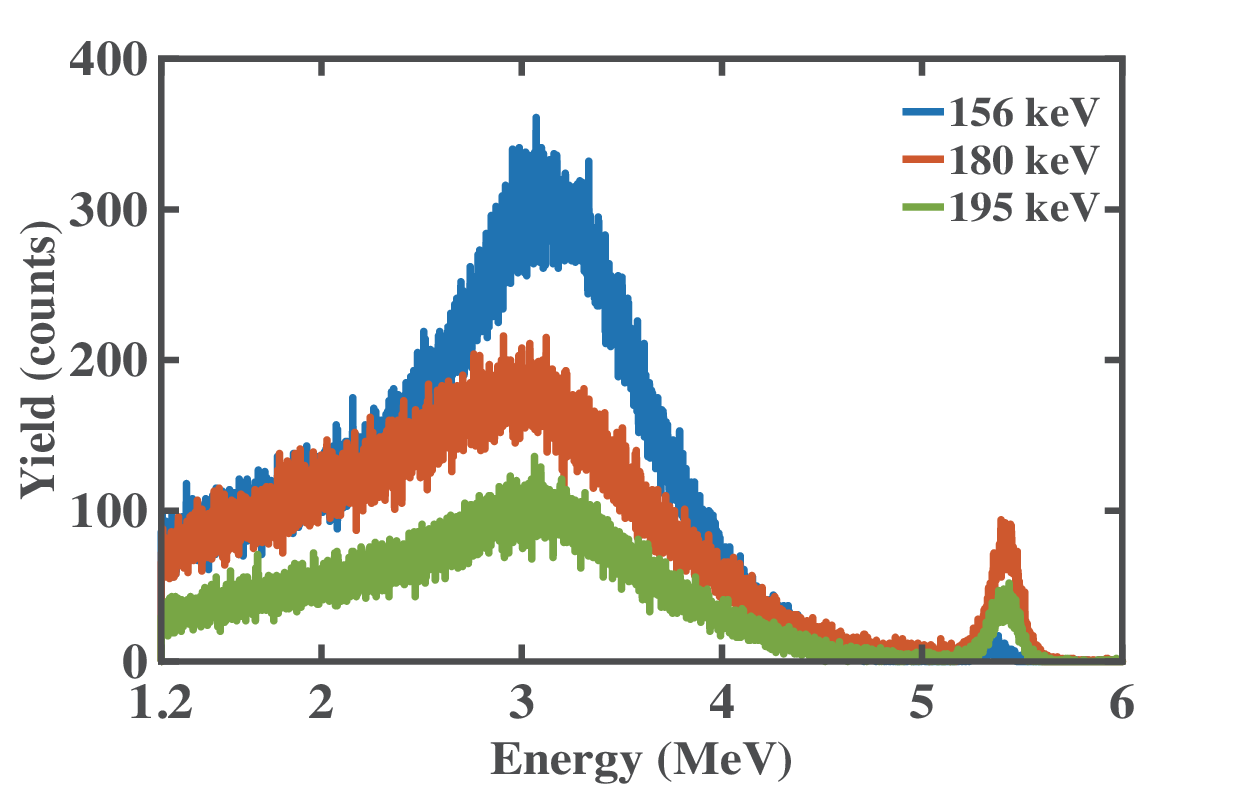}
\caption{Typical $\alpha$-particle energy spectra measured at incident proton energies of
156 keV (blue), 180 keV (red), and 195 keV (green). The $\alpha_0$ peak near 5.4 MeV and
the $\alpha_1$ continuous distribution at lower energies are clearly visible. The entire spectra were shifted to lower energies due to energy loss in the 4\,$\mu$m-thick Mylar film.}
\label{fig2}
\end{figure}

Typical $\alpha$-particle energy spectra are shown in Fig.~\ref{fig2}. The prominent
peak near 5.4 MeV corresponds to $\alpha$ particles from the sequential decay of the
$^{12}$C 16.1 MeV excited state via the $^{8}$Be ground state (GS), denoted as
$\alpha_0$. The broad continuous distribution at lower energies arises from
$\alpha_1$ particles produced by the decay of the first excited state of $^{8}$Be and
associated secondary $\alpha$ components. The $\alpha_0$ and $\alpha_1$ features are clearly
distinguishable, with the $\alpha_1$ peak at approximately 3 MeV and the $\alpha_0$ peak at approximately 5.4 MeV. The entire spectra are shifted toward lower energies due to the energy loss of $\alpha$  particles in the Mylar film. Considering the detector with high noise at low energy range, the data below 1.2\,MeV was cut off in the analysis.

For the measured $\alpha$-particle energy spectra from thick boron target, the $\alpha_1$ particles and the corresponding secondary $\alpha$ components overlap in a broad, continuous distribution, making it impossible to precisely determine the $\alpha_1$ yield by direct peak integration. Previous studies have addressed this difficulty in several ways. Early measurements adopted an empirical division factor to the total $\alpha$ yield, but the choice between factors of 2 and 3 led to the normalization discrepancy, as discussed in the introduction section~\cite{davidson_low_nodate,becker_low-energy_nodate}.
Sikora and Weller~\cite{sikora_new_2016} employed the analytical $\alpha$-energy spectrum model proposed by Stave $\textit{et al.}$~\cite{stave_understanding_2011} to extrapolate the
measured spectrum to zero energy before integration. However, this analytical model provides only the spectral shape, so it still requires the division by a factor of 3 for the total integrated yield to obtain the absolute cross section. Munch and Fynbo~\cite{Munch2018ThePW}
determined the $\alpha$-particle multiplicity directly via DSSD-based kinematic reconstruction, but this technique requires complex multi-detector arrays poorly suited for systematic energy scanning.

Considering these limitations, we employed a Monte Carlo method to simulate the full reaction kinematics. Since the simulation traces each individual reaction event and its subsequent detection, the target-detector configuration is accounted for event by event. It gives how many of the three kinematically correlated $\alpha$ particles lie within the detector acceptance,enabling the measured spectrum to be decomposed into pre-event $\alpha_0$ and $\alpha_1$ components. This therefore eliminates the empirical division factor, and the resulting spectrum carries the detection response of each event rather than the sole spectral shape. Moreover, by simulating the full $\alpha$-energy spectrum down to zero energy, this approach naturally recovers the low-energy $\alpha$ particles that lie below the experimental detection threshold. Aligning the simulated spectrum with the experimental one in the
continuous region above the cutoff, thus the true $\alpha_1$ yield can be directly obtained.

In the simulations, the proton energy was discretized into $n$ intervals, corresponding to different reaction depths within the thick target. The relative reaction probability at each energy was weighted according to Nevins's cross-section parameterization~\cite{nevins_thermonuclear_2000}:
\begin{equation}
P_i = \frac{\sigma(E_i)}{\sum_{j=1}^{n} \sigma(E_j)},
\quad
N_i = N_{\mathrm{total}} \times P_i,
\end{equation}
where $\sigma(E_i)$ is the cross section at proton energy $E_i$ evaluated using Nevins's parameterization, $P_i$ is the relative probability that a reaction occurs in the $i$-th energy interval, $N_i$ is the number of Monte Carlo events assigned to the interval, and $N_{\mathrm{total}}$ is the total number of simulated events. Only the relative energy dependence of $\sigma(E)$ is crucial here, since $P_i$ is normalized by $\sum_j\sigma(E_j)$ such that any global normalization cancels out. The absolute $\alpha_1$ yield is determined separately by matching the simulated spectrum to the experimental data. Hence, Nevins's parameterization, which is based on the measurements of Becker $\textit{et al.}$~\cite{becker_low-energy_nodate}, provides only the spectral shape.

In the center-of-mass frame, the reaction Q-values and level widths are taken from the NACRE compilation~\cite{angulo_compilation_1999} and sampled from a Gaussian distribution. For each simulated event, the reaction channel is randomly selected based on the branching ratios of the ground state ($\alpha_0$) to first excited state ($\alpha_1$) of $^{8}$Be. The subsequent decay of the $^{8}$Be nucleus, whether in the ground state or the first excited state, is simulated as a prompt two-body decay. In the rest frame
of the $^{8}$Be nucleus, the two secondary $\alpha$ particles are emitted isotropically. Their four-momenta are then Lorentz-transformed to the laboratory frame using the velocity of the recoiling $^{8}$Be nucleus, which is obtained from momentum conservation.

After all three $\alpha$ particles are computed, a strict geometric acceptance filter is applied in the simulations. The polar and azimuthal angles of each particle are calculated, but only those with a
polar angle falling within the detector's acceptance range are retained. For these accepted particles, the energy loss through the 4\,$\mu$m Mylar film is calculated event-by-event using SRIM stopping powers~\cite{ziegler_srim_2010}. The particle's energy is then added to the histogram, producing the final simulated spectrum.

\begin{figure}[htbp]
\centering
\includegraphics[width=0.48\textwidth]{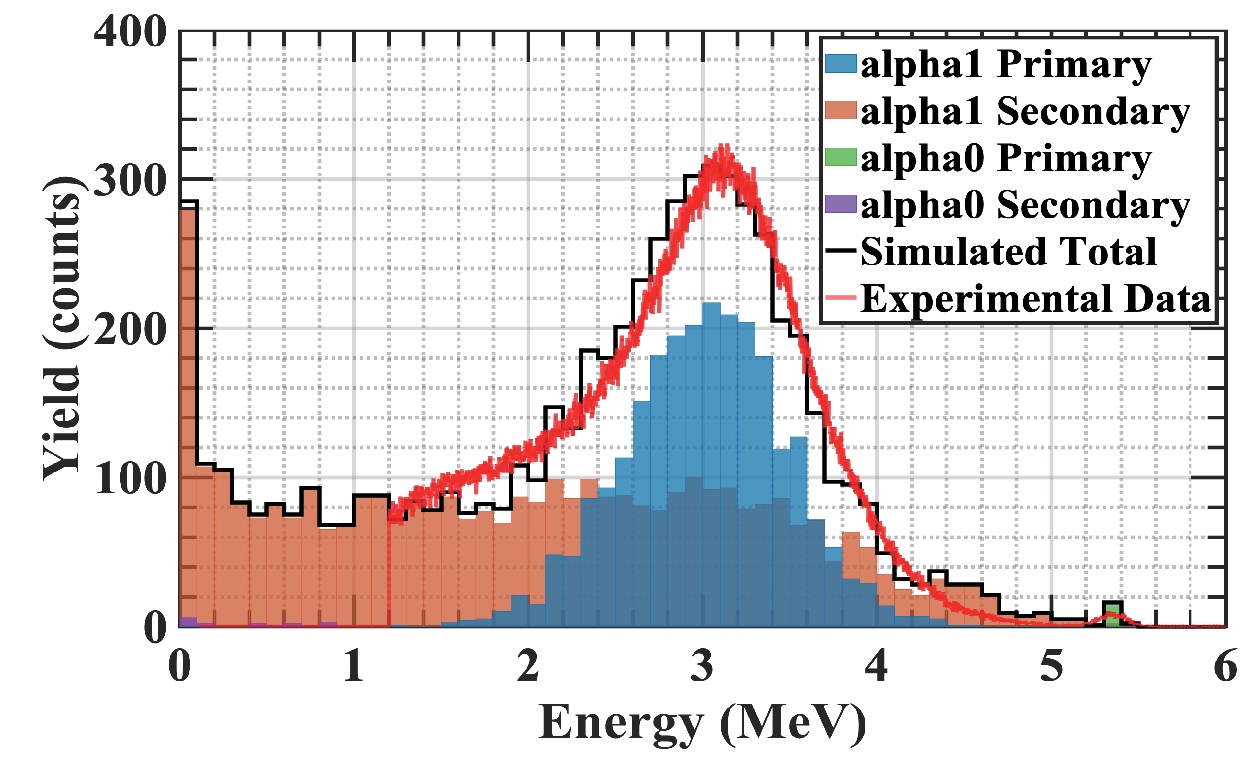}
\caption{Comparison of the Monte Carlo simulated spectrum with the experimental $\alpha$-particle energy spectrum for 156 keV protons. The absolute $\alpha_1$ yield of the simulation is normalized to the experimental data, and the good agreement over the entire energy range validates our simulation method.}
\label{fig3}
\end{figure}

The simulated spectrum is inherently relative, representing the expected
spectral shape for a given count of simulated reaction events. To obtain the absolute $\alpha_1$ yield, the simulated spectrum is scaled to the experimental spectrum in the continuum-dominated region above the 1.2\,MeV threshold, as shown in Fig.~\ref{fig3}. Before matching, the experimental spectrum was smoothed with a five-point moving average to reduce detector noise. The normalization factor is determined by maximizing the coefficient of determination between the simulated and smoothed experimental spectra. The absolute count of $\alpha_1$ events ($N_{\alpha_1}$) is then obtained by integrating the normalized simulated spectrum down to zero energy, thereby recovering all events lost below the experimental threshold. The $\alpha_0$ yield is obtained directly by integrating the experimental peak near 5.3\,MeV. The total reaction count is then as $N_{\mathrm{reaction}} = N_{\alpha_1} + N_{\alpha_0}$.

Once $N_{\mathrm{reaction}}$ is determined, the thick-target reaction probability for each incident proton energy $E_0$ is defined as:
\begin{equation}
P_{\mathrm{thick,exp}}(E_0) = \frac{N_{\mathrm{reaction}}}{N_{\mathrm{p}}},
\end{equation}
where $N_{\mathrm{p}}$ is the total number of incident protons measured by the Faraday cup.
 
The measured probability is related to the reaction cross section by:
\begin{equation}
\begin{aligned}
P_{\mathrm{thick,exp}}(E_0) &= n_{\mathrm{B}} \frac{\Delta\Omega}{4\pi} \int_{0}^{E_{\mathrm{0}}} \frac{\mathrm{d}\Omega_{\mathrm{c.m.}}}{\mathrm{d}\Omega_{\mathrm{lab}}} \\
                   &\quad \times W(E,\theta) \sigma(E) \left(\frac{\mathrm{d}E}{\mathrm{d}x}\right)^{-1} \mathrm{d}E.
\end{aligned}
\label{eq:thick_integral}
\end{equation}
Here, $\Delta\Omega$ is the detector solid angle, $n_{\mathrm{B}}$ is the number density of the $^{11}$B target, $\sigma(E)$ is the cross section at proton energy $E$, $(\mathrm{d}E/\mathrm{d}x)^{-1}$ is the inverse stopping power of protons in $^{11}$B, and $W(E,\theta)$ is the angular distribution function taken from Davidson $\textit{et al.}$~\cite{davidson_low_nodate}.

To extract the cross section from the thick target measured yield, 
we differentiated the yield between adjacent energy points to eliminate the cumulative integration effect.  Data points with unphysical negative derivatives due to statistical fluctuations were also excluded.
From Eq.~\eqref{eq:thick_integral}, the difference between two adjacent incident energies isolates the contribution from the energy interval
$[E_0-\Delta E, \, E_0]$. The thin-target-equivalent yield $P_{\mathrm{thin}}(E_{\mathrm{eff}}) =
P_{\mathrm{thick}}(E_0) - P_{\mathrm{thick}}(E_0 - \Delta E)$ is then related to the cross section as:
\begin{equation}
\begin{aligned}
P_{\mathrm{thin}}(E_{\mathrm{eff}}) &= n_{\mathrm{B}} \frac{\Delta\Omega}{4\pi} \frac{\mathrm{d}\Omega_{\mathrm{c.m.}}}{\mathrm{d}\Omega_{\mathrm{lab}}}
   \int_{E_0-\Delta E}^{E_0} \\
                  &\quad \times W(E,\theta) \, \sigma(E_{\mathrm{eff}})
                     \left(\frac{\mathrm{d}E}{\mathrm{d}x}\right)^{-1} \mathrm{d}E,
\end{aligned}
\label{eq:thin_integral}
\end{equation}
where $E_{\mathrm{eff}}$ is the effective proton energy, taken as the average of $E_0$ and $E_0-\Delta E$. The cross section $\sigma(E_{\mathrm{eff}})$ is treated as constant over the narrow interval.

\begin{figure}[htbp]
\centering
\includegraphics[width=0.48\textwidth]{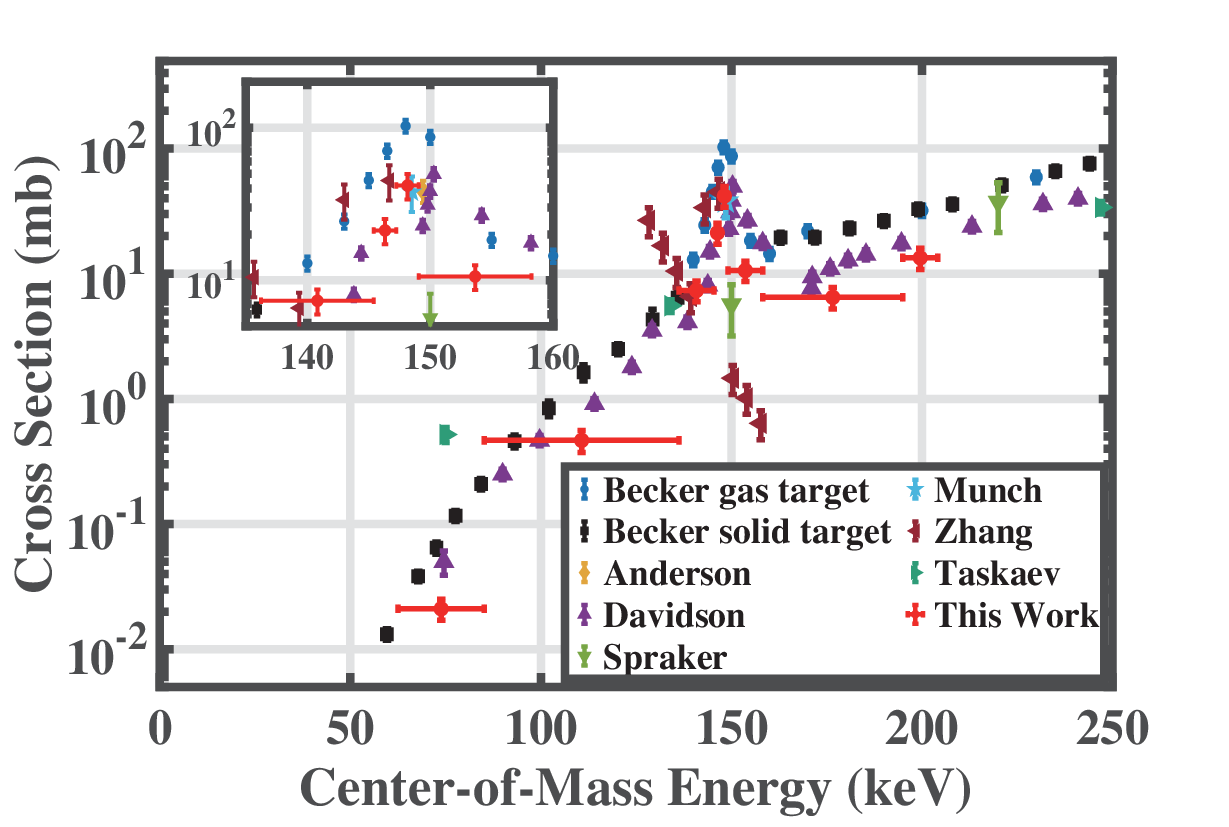}
\caption{Measured cross sections near the first resonance from different research groups. Our thick-target results are in agreement with the Munch $\textit{et al.}$ data and lie below the Becker $\textit{et al.}$values, confirming that the division-by-3 normalization for $\alpha_1$ is correct.}
\label{fig4}
\end{figure}

The resulting cross section is shown in Fig.~\ref{fig4} together with
measurements from other research groups~\cite{becker_low-energy_nodate,spraker_11bp8be_2012,mazzucconi_evaluation_2025,zhang_cross-section_2023,anderson_new_1974,Taskaev_2024}. The cross section reported by Becker $\textit{et al.}$~\cite{becker_low-energy_nodate} is significantly higher than those reported by other groups~\cite{anderson_new_1974,davidson_low_nodate,spraker_11bp8be_2012,Munch2018ThePW,Taskaev_2024,zhang_cross-section_2023}. This discrepancy stems from their adoption of a factor-of-2 division for $\alpha_1$ counting. Our results, derived without any empirical correction factors, show good agreement with the measurements of Munch $\textit{et al.}$~\cite{Munch2018ThePW}. This independently confirms their conclusion that division by 3 is the correct normalization. In addition, this consistency demonstrates the thick-target method developed in our work is effective for extracting cross-section data in the resonance region.

\subsection{Resonance parameter extraction}\label{subseci2}

\begin{figure}[htbp]
\centering
\includegraphics[width=0.48\textwidth]{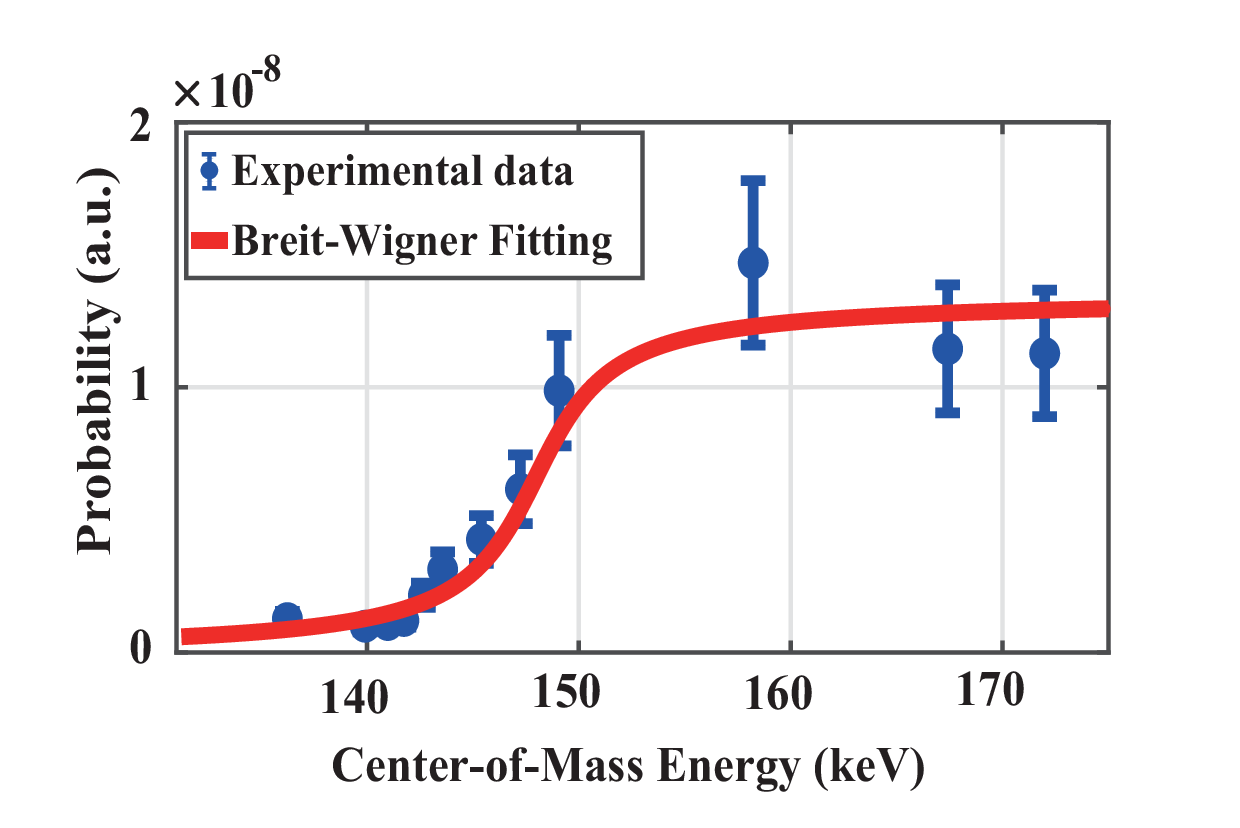}
\caption{Experimental thick-target $\alpha$-particle yield together with the best-fit Breit-Wigner curve in the resonance region, giving  $E_{\mathrm{r}}=148.0$ keV and $\Gamma_{\mathrm{cm}}=5.3$ keV (c.m.).}
\label{fig5}
\end{figure}

The thick-target method for cross-section extraction has been validated in Sec.~\ref{subseci1}. Beside the cross section, resonance parameters as important physical quantities are also essential for evaluating the reaction rate and can be used to characterize the excited state of $^{12}$C. These parameters can likewise be obtained from the thick-target measured yield data. The thick-target reaction probability can be described by the Breit-Wigner formula~\cite{Munch2018ThePW} in the center-of-mass frame as:
{\setlength\abovedisplayskip{2pt}\setlength\belowdisplayskip{2pt}
\begin{equation}
\begin{aligned}
P(E_0) &= \biggl[ \tan^{-1}\!\biggl(\frac{E_{\mathrm{0}} - E_{\mathrm{r}}}{\Gamma_{\mathrm{cm}}/2}\biggr) \\
       &\qquad - \tan^{-1}\!\biggl(\frac{E_{\mathrm{0}} - E_{\mathrm{r}} - \Delta E}{\Gamma_{\mathrm{cm}}/2}\biggr) \biggr] \\
       &\quad \times \frac{2\pi \Gamma_{\mathrm{cm}} \sigma_{\mathrm{BW}} n_{\mathrm{B}}}{N_{\mathrm{p}} \Delta\Omega \left(\frac{\mathrm{d}E}{\mathrm{d}x}\right)^{-1}},
\end{aligned}
\end{equation}}
where $\Gamma_{\mathrm{cm}}$ is the resonance level width, $E_{\mathrm{0}}$ the incident beam energy, $E_{\mathrm{r}}$ the resonance energy, $\Delta E$ the energy loss of protons in the thick target (here $\Delta E = E_{\mathrm{0}}$), $\sigma_{\mathrm{BW}}$ the resonant
Breit-Wigner cross section. Treating the factor outside the brackets as a normalization constant, the best fit gave $E_{\mathrm{r}} = 148.0$ keV (c.m.) and $\Gamma_{\mathrm{cm}} = 5.3$ keV. The experimental yield data and the Breit-Wigner fit curve are shown in Fig.~\ref{fig5}. The results demonstrate that the Breit-Wigner form reproduces the data within experimental uncertainties. The obtained resonance parameters agree with the values from thin-target measurements, as summarized in Table~\ref{tab1}.

\begin{table}[h]
\caption{Resonance parameters obtained by different research groups. The present values are obtained from the Breit-Wigner fit of the measured yield.}\label{tab1}
\begin{tabular}{lcc}
\hline
Research Group & $E_{\mathrm{r}}$ (keV, c.m.) & $\Gamma_{\mathrm{cm}}$ (keV, c.m.) \\
\hline
This work & 148.0 & 5.3 \\
Munch $\textit{et al.}$~\cite{Munch2018ThePW} & 149.05 & 5.28 \\
Becker $\textit{et al.}$~\cite{becker_low-energy_nodate} & 148.3 & 5.3 \\
Davidson $\textit{et al.}$~\cite{davidson_low_nodate} & 149.8 & 5.2 \\
\hline
\end{tabular}
\end{table}

\subsection{Low-energy resonance search}\label{subseci3}

Building on the validated thick-target method, we extend this approach to search for low-energy resonances, where the high beam current compensates for the extremely small cross section in the ultra-low-energy region. The search proceeds by comparing the measured thick-target yield with a theoretical extrapolation data that assumes no additional resonance. Specifically, the cross section data for the first resonance peak are fitted using the following parameterization:
\begin{equation}
\sigma(E) = A \, \frac{S(E)}{E} \exp\left(-\sqrt{\frac{E_{\mathrm{G}}}{E}}\right),
\end{equation}
where $E_{\mathrm{G}} = 22.589$ MeV is the Gamow energy for the p$^{11}$B reaction, $S(E)$ is the S-factor derived from Nevins and Swain~\cite{nevins_thermonuclear_2000}, and $A$ is a normalization coefficient adjusted to fit the experimental data. The fitted parameters are then used to extrapolate the cross section to lower energies, assuming no additional low-energy resonance, yielding $\sigma_{\mathrm{theory}}(E)$. Hence, the expected thick-target reaction probability $P_{\mathrm{theory}}(E)$ is obtained via integration of $\sigma_{\mathrm{theory}}(E)$. Meanwhile, the experimental thick-target reaction probability $P_{\mathrm{exp}}(E)$ is measured at a series of selected low-energy points. A systematic excess of $P_{\mathrm{exp}}(E)$ over $P_{\mathrm{theory}}(E)$ would indicate the possible existence of a low-energy resonance, whereas consistency between the two quantities would imply no significant resonance structure within the probed energy region.

\begin{figure}[htbp]
\centering
\includegraphics[width=0.48\textwidth]{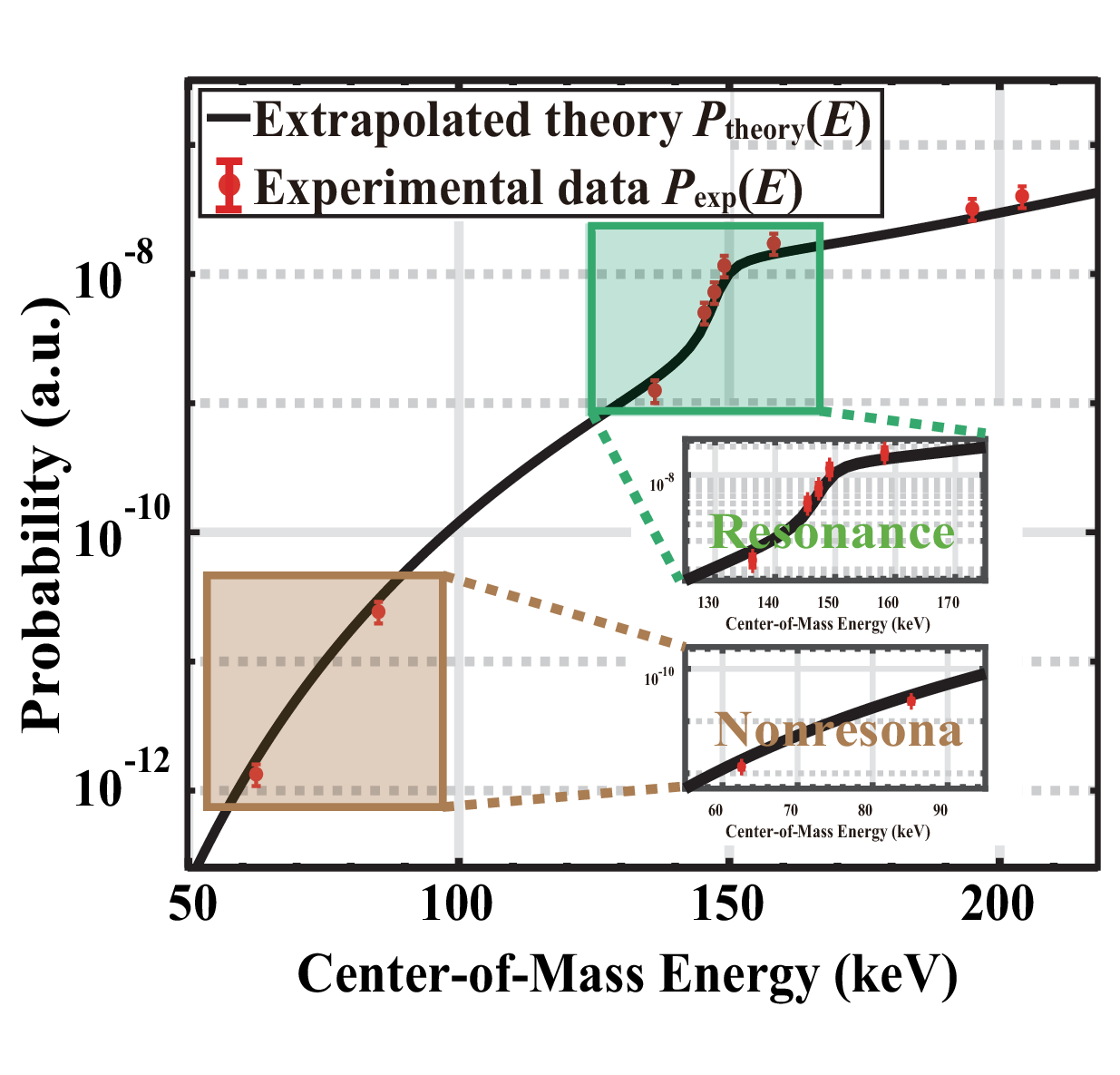}
\caption{Thick-target method given reaction probability curve with two magnified insets. The upper inset shows the resonance region around 148.0\,keV, where a steep rising edge marks the characteristic of a resonance. The lower inset shows measurements at proton energies of  75\,keV and 100\,keV in the non-resonance region, consistent with the no-resonance extrapolation.}
\label{fig6}
\end{figure}

Figure~\ref{fig6} shows the reaction probability profile obtained via thick-target method, with two magnified insets in the lower-right corner. The upper inset focuses on the region around the known 148.0\,keV resonance, where a steep rise is clearly visible. This sharp rise constitutes the characteristic signature of a resonance in thick-target yield and serves as a sensitive marker for identifying resonance structures in the low-energy regime. The lower inset displays the non-resonance region, where measurements at proton energies of 75\,keV and 100\,keV are consistent with the no-resonance extrapolation within
uncertainties, revealing no abrupt spectral features.

This approach can serve as an efficient scanning tool. By surveying a broad low-energy range, it can identify energy regions where cross-section enhancement may occur, guiding subsequent high-resolution
thin-target measurements or indirect spectroscopic experiments. If a
deviation is detected, the Monte Carlo inversion and Breit-Wigner fitting methods developed in this work can be directly applied to extract the resonance energy, width, and cross section. This methodology
can also be extended to search for low-energy resonances in other
nuclear reactions for which low-energy cross-section data are scarce.

\section{Conclusions}\label{sec4}
A thick-target methodology was developed for the
$^{11}$B(p,$\alpha$)2$\alpha$ reaction to measure the cross section and resonance parameters. Utilizing beam currents of tens of microamperes, which are orders of magnitude higher than the nanoampere limit of conventional thin-target measurements, this method compensates for the extremely low cross section in the ultra-low-energy regime. Monte Carlo simulations of the reaction kinematics were used to determine the reaction yield by matching the simulated and experimental $\alpha$-particle spectra, without relying on empirical division factors. Breit-Wigner fitting of the thick-target yield curve gave a resonance energy of $E_{\mathrm{r}} = 148.0$ keV (c.m.) and a level width of $\Gamma_{\mathrm{cm}} = 5.3$ keV, in good agreement with published thin-target measurements. This demonstrates the capability of the thick-target approach for extracting both cross section and resonance parameters.

Based on these results, we demonstrate a scheme to probe low-energy resonances within the ultra-low-energy regime. It compares the measured thick-target reaction probability with a theoretical extrapolation from the first resonance peak under the assumption of no additional low-energy resonance. Preliminary measurements at proton energies of 75 keV and 100 keV reveal no evidence for a low-energy resonance. The scheme provides an efficient scanning tool for systematic searches at ultra-low energies, and can be extended to other nuclear reactions where low-energy cross-section data are scarce. Future work will extend the measurements to lower energies with finer steps, aiming to survey the entire low-energy region below 100\,keV.

\acks

We sincerely thank the staff of the high-power proton linear accelerator facility at the Institute of Modern Physics, Chinese Academy of Sciences, for operating the ion-beam system. This work has been supported by the National Key Research and Development Program of China (Grant No.~2022YFA1603302), Chinese Science Challenge Project (Grant No.~TZ2025012), National Natural Science Foundation of China (Grant Nos.~12105218, 12422512, U2541245, 12595365 and 12120101005), China Postdoctoral Science Foundation (Grant No.~2020M673367), ENN's Hydrogen-Boron Fusion Research Fund (Grant No.~2025ENNHB01-004), Key Research and Development Program of Shaanxi Province (Grant No.~2024PT-ZCK-83), and Innovative Scientific Program of CNNC.

\bibliographystyle{unsrtnat}
\bibliography{sn-bibliography}

@article{Moreau_1977,
url = {https://doi.org/10.1088/0029-5515/17/1/002},
year = {1977},
month = {feb},
publisher = {},
volume = {17},
number = {1},
pages = {13},
author = {Moreau, D.C.},
title = {Potentiality of the proton-boron fuel for controlled thermonuclear fusion},
journal = {Nuclear Fusion},
}

@Article{app12031444,
AUTHOR = {Margarone, Daniele and Bonvalet, Julien and Giuffrida, Lorenzo and Morace, Alessio and Kantarelou, Vasiliki and Tosca, Marco and Raffestin, Didier and Nicolai, Philippe and Picciotto, Antonino and Abe, Yuki and Arikawa, Yasunobu and Fujioka, Shinsuke and Fukuda, Yuji and Kuramitsu, Yasuhiro and Habara, Hideaki and Batani, Dimitri},
TITLE = {In-Target Proton–Boron Nuclear Fusion Using a {PW}-Class Laser},
JOURNAL = {Applied Sciences},
VOLUME = {12},
YEAR = {2022},
NUMBER = {3},
ARTICLE-NUMBER = {1444},
URL = {https://www.mdpi.com/2076-3417/12/3/1444},
ISSN = {2076-3417},
}

@article{Valeria2023,
title = "A multi-{MeV} alpha particle source via protonboron fusion driven by a 10-{GW} tabletop laser",
author = "Valeria Istokskaia and Marco Tosca and L. Giuffrida and J. Psikal and Vasiliki Kantarelou and Filip Grepl and Stanislav Stancek and Di Siena, Sabrina and Arsenios Hadjikyriacou and Aodhan McIlvenny and Yoann Levy and Jaroslav Huynh and Martin Cimrman and Pavel Pleskunov and Daniil Nikitin and A. Choukourov and Fabio Belloni and Antonino Picciotto and Satyabrata Kar and Marco Borghesi and Antonio Lucianetti and Tomas Mocek and Daniele Margarone",
year = "2023",
month = feb,
day = "2",
language = "English",
volume = "6",
journal = "Communications Physics",
issn = "2399-3650",
publisher = "Springer Nature US",
}

@article{Batani_Margarone_Belloni_2023, 
title={Advances in the Study of Laser-Driven Proton-Boron Fusion}, volume={2023}, 
journal={Laser and Particle Beams}, 
author={Batani, Dimitri and Margarone, Daniele and Belloni, Fabio}, 
year={2023}, 
pages={e5}}

@article{PhysRevE.106.055215,
  title = {Improving the feasibility of economical proton-boron-11 fusion via alpha channeling with a hybrid fast and thermal proton scheme},
  author = {Ochs, Ian E. and Kolmes, Elijah J. and Mlodik, Mikhail E. and Rubin, Tal and Fisch, Nathaniel J.},
  journal = {Phys. Rev. E},
  volume = {106},
  issue = {5},
  pages = {055215},
  numpages = {14},
  year = {2022},
  month = {Nov},
  publisher = {American Physical Society},
  url = {https://link.aps.org/doi/10.1103/PhysRevE.106.055215}
}

@article{Ogawa_2024,
url = {https://doi.org/10.1088/1741-4326/ad6615},
year = {2024},
month = {aug},
publisher = {IOP Publishing},
volume = {64},
number = {9},
pages = {096028},
author = {Ogawa, K. and Magee, R.M. and Tajima, T. and Gota, H. and McCarroll, P. and Allfrey, I. and Nuga, H. and Isobe, M. and Osakabe, M.},
title = {Demonstration of aneutronic p-11{B} reaction in a magnetic confinement device},
journal = {Nuclear Fusion}
}

@Article{Magee2023,
journal={Nature Communications},
author={R. M. Magee and K. Ogawa and T. Tajima and I. Allfrey and H. Gota and P. McCarroll and S. Ohdachi and M. Isobe and S. Kamio and V. Klumper and H. Nuga and M. Shoji and S. Ziaei and M. W. Binderbauer and M. Osakabe},
title={First measurements of p11{B} fusion in a magnetically confined plasma},
year={2023},
month={December},
pages={1-6},
volume={14},
number={1},
url={https://ideas.repec.org/a/nat/natcom/v14y2023i1d10.1038_s41467-023-36655-1.html},
}

@article{li2024proton,
  title={Proton--boron fusion in a hydrogen-doped-boron target},
  author={Li, Zhi and Wang, Zhao and Zhao, Guanchao and Liu, Bing and Xie, Huasheng and Tan, Muzhi and Huang, Hairong and Liu, Minsheng and Hoffmann, Dieter HH and Cheng, Rui and others},
  journal={Laser and Particle Beams},
  volume={42},
  pages={e5},
  year={2024},
  publisher={Cambridge University Press}
}

@article{Labaune2013FusionRI,
  title={Fusion reactions initiated by laser-accelerated particle beams in a laser-produced plasma},
  author={Christine Labaune and C. Baccou and S. Depierreux and Clement Goyon and G P Loisel and Vincent Yahia and Johann Rafelski},
  journal={Nature Communications},
  year={2013},
  volume={4},
  url={https://api.semanticscholar.org/CorpusID:7802026}
}

@article{giuffrida_high-current_2020,
	title = {High-current stream of energetic $\alpha$ particles from laser-driven proton-boron fusion},
	volume = {16},
	issn = {2041-1723},
	language = {en},
	number = {11},
	journal = {Nature Communications},
	author = {Giuffrida, L. and Margarone, D. and and others},
	month = nov,
	year = {2020},
	pages = {1--10},
	nameformat={lowercase}
}

@article{bonvalet_energetic_2021,
	title = {Energetic $\alpha$-particle sources produced through proton-boron reactions by high-energy high-intensity laser beams},
	volume = {29},
	issn = {1070-664X},
	language = {en},
	number = {12},
	journal = {Physics of Plasmas},
	author = {Bonvalet, J. and and others},
	month = dec,
	year = {2021},
	pages = {123101},
	nameformat={lowercase}
}

@article{picciotto_boron-proton_2014,
	title = {Boron-proton nuclear-fusion enhancement induced by boron-doped targets in femtosecond laser plasmas},
	volume = {72},
	language = {en},
	journal = {Physical Review E},
	author = {Picciotto, A. and Margarone, D. and and others},
	year = {2014},
	pages = {046402},
	nameformat={lowercase}
}

@article{luo2026gated,
  author = {Luo, Xu-Yang and Wei, Wen-Qing and Ren, Jie-Ru and Zhang, Shi-Zheng and Ma, Bu-Bo and Zheng, Bo-Wei and Zhao, Zi-Qian and Cui, Bo and Qi, Wei and Deng, Zhi-Gang and Shi, Lu-Lin and Cheng, Rui and Li, Kui-Nian and Chen, Ping and Hoffmann, Dieter H.H. and Zhao, Yong-Tao},
  title = {A gated alpha-particle time-of-flight spectrometer for laser-driven proton-boron fusion reactions},
  journal = {Review of Scientific Instruments},
  volume = {97},
  number = {7},
  pages = {073504},
  year = {2026},
  month = jul,
  doi = {10.1063/5.0341327}
}

@article{sciscio2025laser,
  author = {Scisciò, M. and Petringa, G. and Zhu, Z. and Rodrigues, M. R. D. and Alonzo, M. and Andreoli, P. L. and Filippi, F. and Consoli, Fe. and Huault, M. and Raffestin, D. and Molloy, D. and Larreur, H. and Singappuli, D. and Carriere, T. and Verona, C. and Nicolai, P. and McNamee, A. and Ehret, M. and Filippov, E. and Lera, R. and Pérez-Hernández, J. A. and Agarwal, S. and Krupka, M. and Singh, S. and Istokskaia, V. and Lattuada, D. and La Cognata, M. and Guardo, G. L. and Palmerini, S. and Rapisarda, G. and Batani, K. and Cipriani, M. and Cristofari, G. and Di Ferdinando, E. and Di Giorgio, G. and De Angelis, R. and Giulietti, D. and Xu, J. and Volpe, L. and Rodríguez-Frías, M. D. and Giuffrida, L. and Margarone, D. and Batani, D. and Cirrone, G. A. P. and Bonasera, A. and Consoli, Fa.},
  title = {Laser-initiated p–11{B} fusion reactions in petawatt high-repetition-rate laser facilities},
  journal = {Matter and Radiation at Extremes},
  volume = {10},
  number = {3},
  pages = {037402},
  year = {2025},
  month = apr,
  doi = {10.1063/5.0241993}
}

@article{Jelinek2024,
  title={First independent validation of the proton-boron capture therapy concept},
  author={Jel{\'\i}nek Michaelidesov{\'a}, Anna and Kundr{\'a}t, Pavel and Zahradn{\'\i}{\v{c}}ek, Old{\v{r}}ich and Danilov{\'a}, Irina and Pachnerov{\'a} Brabcov{\'a}, Kate{\v{r}}ina and Vachelov{\'a}, Jana and Vilimovsk{\`y}, Jan and David, Miroslav and Vondr{\'a}{\v{c}}ek, Vladim{\'\i}r and Dav{\'\i}dkov{\'a}, Marie},
  journal={Scientific Reports},
  volume={14},
  number={1},
  pages={19264},
  year={2024},
  publisher={Nature Publishing Group UK London}
}

@article{Yoon2014,
    author = {Yoon, Do-Kun and Jung, Joo-Young and Suh, Tae Suk},
    title = {Application of proton boron fusion reaction to radiation therapy: A {Monte Carlo} simulation study},
    journal = {Applied Physics Letters},
    volume = {105},
    number = {22},
    pages = {223507},
    year = {2014},
    month = {12},
    issn = {0003-6951},
    url = {https://doi.org/10.1063/1.4903345}
}

@article{Tran2023CurrentSA,
  title={Current State and Prospectives for Proton Boron Capture Therapy},
  author={Nhan Hau Tran and Tatiana A. Shtam and Yaroslav Marchenko and Andrey L. Konevega and Dmitry V. Lebedev},
  journal={Biomedicines},
  year={2023},
  volume={11},
  url={https://api.semanticscholar.org/CorpusID:259269306}
}

@article{Mckenzie2023,
title={{HB11}---understanding hydrogen-boron fusion as a new clean energy source},
  author={McKenzie, Warren and Batani, Dimitri and Mehlhorn, Thomas A and Margarone, Daniele and Belloni, Fabio and Campbell, E Michael and Woodruff, Simon and Kirchhoff, Jan and Paterson, Adrian and Pikuz, Sergey and others},
  journal={Journal of Fusion Energy},
  volume={42},
  number={1},
  pages={17},
  year={2023},
  publisher={Springer}
}

@article{Laursen2016,
title = "Complete kinematical study of the 3 alpha breakup of the 16.11 {MeV} state in {C-12}",
author = "Laursen, \{K. L.\} and Fynbo, \{H. O. U.\} and Kirsebom, \{O. S.\} and Madsboll, \{K. S.\} and K. Riisager",
year = "2016",
month = sep,
day = "1",
doi = "10.1140/epja/i2016-16271-2",
language = "English",
volume = "52",
journal = "The European Physical Journal A: Hadrons and Nuclei",
issn = "1434-6001",
publisher = "Springer",
number = "9",
}

@article{kelley_energy_2017,
	title = {Energy levels of light nuclei {A}=12},
	volume = {968},
	issn = {03759474},
	language = {en},
	journal = {Nuclear Physics A},
	author = {Kelley, J. H. and Purcell, J. E. and Sheu, C. G.},
	month = dec,
	year = {2017},
	pages = {71--253},
	nameformat={lowercase}
}

@article{Boesgaard_2005,
url = {https://doi.org/10.1086/427687},
year = {2005},
month = {mar},
publisher = {},
volume = {621},
number = {2},
pages = {991},
author = {Boesgaard, Ann Merchant and Deliyannis, Constantine P. and Steinhauer, Aaron},
title = {Boron Depletion in {F} and {G} Dwarf Stars and the Beryllium-Boron Correlation*},
journal = {The Astrophysical Journal}
}

@article{oliphant_experiemnts_1933,
	author = {Oliphant, Marcus Laurence Elwin and Rutherford, Ernest},
	title = {Experiments on the transmutation of elements by protons},
	journal = {Proceedings of the Royal Society of London. Series A, Containing Papers of a Mathematical and Physical Character},
	volume = {141},
	number = {843},
	pages = {259-281},
	year = {1933},
	nameformat={lowercase}
}

@article{dee_disintergratin_1936,
	author = {Dee, Philip Ivor and Gilbert, C. W. and Rutherford, Ernest},
	title = {The disintegration of boron into three $\alpha$-particles},
	journal = {Proceedings of the Royal Society of London. Series A - Mathematical and Physical Sciences},
	volume = {154},
	number = {881},
	pages = {279-296},
	year = {1936},
	nameformat={lowercase}
}

@article{davidson_low_nodate,
	title = {Low energy cross section for $^{11}${B}(p,3$\alpha$)},
	language = {en},
	author = {Davidson, J. M. and Berg, H. L. and Lowry, M. M.},
	journal = {Nuclear Physics},
	volume = {A315},
	pages = {253-268},
	year = {1979},
	nameformat={lowercase}
}

@article{becker_low-energy_nodate,
	title = {Low-energy cross sections for $^{11}${B}(p,3$\alpha$)},
	language = {en},
	journal = {Zeitschrift Fur Physik A - Hadrons and Nuclei},
	year = {1987},
	volume = {327},
	number = {3},
	pages = {341-355},
	author = {Becker, H. W. and Rolfs, C. and Trautvetter, H. P.},
	nameformat={lowercase}
}

@article{angulo_compilation_1999,
	title = {A compilation of charged-particle induced thermonuclear reaction rates},
	language = {en},
	journal = {Nuclear Physics A},
	author = {Angulo, C. and Arnould, M. and Rayet, M. and Descouvemont, P. and Baye, D. and Leclercq-Willain, C. and Coc, A. and Barhoumi, S. and Aguer, P. and Rolfs, C. and Kunz, R. and Hammer, J. W. and Mayer, A. and Paradellis, T. and Kossionides, S. and Chronidou, C. and Spyrou, K. and Degl'Innocenti, S. and Fiorentini, G. and Ricci, B. and Zavatarelli, S. and Providencia, C. and Wolters, H. and Soares, J. and Grama, C. and Rahighi, J. and Shotter, A. and Lamehi Rachti, M.},
	month = aug,
	year = {1999},
	pages = {3--183},
	nameformat={lowercase}
}

@article{nevins_thermonuclear_2000,
	title = {The thermonuclear fusion rate coefficient for p$^{11}${B} reactions},
	language = {en},
	volume = {40},
	number = {4},
	journal = {Nuclear Fusion},
	author = {Nevins, W. M. and Swain, R.},
	month = apr,
	year = {2000},
	pages = {865--872},
	nameformat={lowercase}
}

@article{spraker_11bp8be_2012,
	title = {The $^{11}${B}(p,$\alpha_0$)$^{8}${Be} $\rightarrow$ $\alpha$ + $\alpha$ and the $^{11}${B}($\alpha$,$\alpha$)$^{11}${B} reactions at energies below 5.4 {MeV}},
	volume = {31},
	number = {4},
	journal = {Journal of Fusion Energy},
	author = {Spraker, M. C. and Ahmed, M. W. and Blackston, M. A. and Brown, N. and France, R. H. and Henshaw, S. S. and Perdue, B. A. and Prior, R. M. and Seo, P.-N. and Stave, S. and Weller, H. R.},
	month = aug,
	year = {2012},
	pages = {357--367},
	nameformat={lowercase}
}

@article{sikora_new_2016,
	title = {A New Evaluation of the $^{11}${B}(p,$\alpha$)$\alpha$$\alpha$ Reaction Rates},
	volume = {35},
	issn = {0164-0313, 1572-9591},
	language = {en},
	number = {3},
	journal = {Journal of Fusion Energy},
	author = {Sikora, M. H. and Weller, H. R.},
	month = jun,
	year = {2016},
	pages = {538--543},
	nameformat={lowercase}
}

@article{stave_understanding_2011,
	title = {Understanding the $^{11}${B}(p,$\alpha$)$\alpha$ reaction at the 0.675 {MeV} resonance},
	volume = {696},
	issn = {03702693},
	language = {en},
	number = {1-2},
	journal = {Physics Letters B},
	author = {Stave, S. and Ahmed, M. W. and France, R. H. and Henshaw, S. S. and Muller, B. and Perdue, B. A. and Prior, R. M. and Spraker, M. C. and Weller, H. R.},
	month = jan,
	year = {2011},
	pages = {26--29},
	nameformat={lowercase}
}

@article{Munch2018ThePW,
  title={The partial widths of the 16.1 {MeV} 2+ resonance in 12{C}},
  author={Michael Munch and Hans Otto Uldall Fynbo},
  journal={The European Physical Journal A},
  year={2018},
  volume={54},
  number={8},
  pages={138},
  doi={10.1140/epja/i2018-12577-3},
  url={https://api.semanticscholar.org/CorpusID:254109228}
}

@article{mazzucconi_evaluation_2025,
	title = {Evaluation of the p-$^{11}${B} reaction cross section through a silicon telescope in the 0.3--4.7 {MeV} range},
	volume = {61},
	issn = {1434-601X},
	number = {5},
	journal = {The European Physical Journal A},
	author = {Mazzucconi, D. and Bellotti, E. S. and Vavassori, D. and Dellasega, D. and Agosteo, S. and Passoni, M. and Pola, A. and Bortot, D.},
	month = may,
	year = {2025},
	pages = {114},
	nameformat={lowercase}
}

@article{buck_alpha-cluster_nodate,
	title = {Alpha-cluster break-up and reaction mechanism in (p,$\alpha$) reactions on light nuclei},
	language = {en},
	journal = {Nuclear Physics A},
	year = {1983},
	volume = {398},
	number = {2},
	pages = {189-202},
	author = {Buck, W. and Hoyler, F. and Stabler, A. and Staudt, G. and Klapdor, H. V. and Oeschler, H.},
	nameformat={lowercase}
}

@article{segel_states_1965,
	title = {States in $^{12}${C} between 16.4 and 19.6 {MeV}},
	volume = {139},
	issn = {0031-899X},
	language = {en},
	number = {4B},
	journal = {Physical Review},
	author = {Segel, R. E. and Hanna, S. S. and Allas, R. G.},
	month = aug,
	year = {1965},
	pages = {B818--B830},
	nameformat={lowercase}
}

@article{beckman_excitation_1953,
	title = {Excitation curves for $\alpha$ particles from $^{11}${B} bombarded with protons},
	volume = {91},
	issn = {0031-899X},
	language = {en},
	number = {3},
	journal = {Physical Review},
	author = {Beckman, Olof and Huus, Torben and Zupancic, Crtomir},
	month = aug,
	year = {1953},
	pages = {606--609},
	nameformat={lowercase}
}

@article{anderson_new_1974,
	title = {A new determination of the partial widths of the 16.11 {MeV} state in $^{12}${C}},
	volume = {233},
	issn = {03759474},
	language = {en},
	number = {2},
	journal = {Nuclear Physics A},
	author = {Anderson, B. D. and Dwarakanath, M. R. and Schweitzer, J. S. and Nero, A. V.},
	month = nov,
	year = {1974},
	pages = {286--296},
	nameformat={lowercase}
}

@article{Taskaev_2024,
author = {Taskaev, Sergey and Bessmeltsev, Victor and Bikchurina, Marina and Bykov, Timofey and Kasatov, Dmitry and Kolesnikov, Yaroslav and Nikolaev, Alexey and Oks, E. and Ostreinov, George and Savinov, Sergey and Shuklina, Anna and Sokolova, Evgeniia and Yushkov, Georgy},
year = {2024},
month = {10},
pages = {165490},
title = {Measurement of the 11{B}(p,$\alpha_0$)8{Be} and the 11{B}(p,$\alpha_1$)8{Be}* reactions cross-sections at the proton energies up to 2.2 {MeV}},
volume = {555},
journal = {Nuclear Instruments and Methods in Physics Research Section B: Beam Interactions with Materials and Atoms},
doi = {10.1016/j.nimb.2024.165490}
}

@article{ziegler_srim_2010,
	title = {{SRIM} -- {The} stopping and range of ions in matter (2010)},
	volume = {268},
	issn = {0168583X},
	language = {en},
	number = {11-12},
	journal = {Nuclear Instruments and Methods in Physics Research Section B: Beam Interactions with Materials and Atoms},
	author = {Ziegler, James F. and Ziegler, M. D. and Biersack, J. P.},
	month = jun,
	year = {2010},
	pages = {1818--1823},
	nameformat={lowercase}
}

@article{zhang_cross-section_2023,
	author = {Zhang, Shizheng and Xu, Hao and Xu, Xing and Wei, Wenqing and Ren, Jieru and Chen, Benzheng and Ma, Bubo and Hu, Zhongmin and Li, Fangfang and Liu, Lirong and Yang, Mingzhe and Lai, Zeyu and Yue, Hongwei and Xiong, Jie and Xu, Zhongfeng and Chen, Yan and Wang, Zhao and Zhou, Zexian and Shi, Lulin and Zhao, Yongtao},
	year = {2023},
	month = {02},
	title = {Cross-Section Measurements of the $^{11}${B}(p,$\alpha$)2$\alpha$ Reaction near the First Resonant Energy},
	volume = {2023},
	journal = {Laser and Particle Beams},
	doi = {10.1155/2023/9697329}
}

@article{BROGGINI201855,
	title = {{LUNA}: Status and prospects},
	journal = {Progress in Particle and Nuclear Physics},
	volume = {98},
	pages = {55--84},
	year = {2018},
	doi = {10.1016/j.ppnp.2017.09.002},
	author = {C. Broggini and D. Bemmerer and A. Caciolli and D. Trezzi}
}

@article{zhang2022measurement,
	title = {Measurement of ${}^{19}${F}(p,$\gamma$)${}^{20}${Ne} reaction suggests {CNO} breakout in first stars},
	author = {Zhang, Liyong and He, Jianjun and DeBoer, Richard J. and Wiescher, Michael and Heger, Alexander and Kahl, Daid and Su, Jun and Odell, Daniel and Chen, Yinji and Li, Xinyue and others},
	journal = {Nature},
	volume = {610},
	number = {7933},
	pages = {656--660},
	year = {2022}
}

@article{Tentori_2023,
	doi = {10.1088/1741-4326/acda4b},
	year = {2023},
	volume = {63},
	number = {8},
	pages = {086001},
	author = {Tentori, Alessandro and Belloni, Fabio},
	title = {Revisiting p-{11B} fusion cross section and reactivity, and their analytic approximations},
	journal = {Nuclear Fusion}
}

@book{iliadis_2007,
	title = {Nuclear Physics of Stars},
	author = {Iliadis, Christian},
	year = {2007},
	publisher = {Wiley-VCH},
	address = {Weinheim},
	url = {https://api.semanticscholar.org/CorpusID:117506721}
}

@article{2026LuoDi,
	author = {Luo, Di and Xiao, Xiao and Zhao, Guanchao and Wang, Guodong and Dong, Junyu and Shi, Lulin and Zhang, Yanshi and Cheng, Rui and Liu, Bing and Dong, Jiaqi and Peng, Yuankai and Hoffmann, Dieter H.H. and Liu, Minsheng},
	year = {2026},
	title = {Plasma enhancement of proton-boron fusion via high-current beam-target interaction},
	volume = {44},
	pages = {40004},
	journal = {Laser and Particle Beams},
	doi = {10.3788/LPB.2026.40004}
}

\end{document}